\documentclass[prb,twocolumn,showpacs]{revtex4}
\usepackage{graphics}
\usepackage{epsfig}

\begin{document}

\title{Atomistic, microstructural and micromagnetic aspects of
the multiscale modelling of hysteretic phenomena}
\author{K. D. Belashchenko}
\author{V. P. Antropov}
\affiliation{Ames Laboratory, Ames, Iowa 50011}

\begin{abstract}
We formulated a technique which combines the first-principles,
micromagnetic and microstructural calculations and allows us to
study the nature of hysteretic phenomena in hard magnets.
Two distinct sources of coercivity in polytwinned CoPt type
magnets, domain wall pinning at antiphase boundaries and
splitting at twin boundaries, are illustrated for
a realistic microstructure.
Methodology of multiscale modelling of hysteretic phenomena
in nanoscale magnets is discussed.
\end{abstract}

\maketitle

\section{Introduction}

It is a well established experimental fact that the hystetic
properties of a magnetic material are very sensetive both to
its magnetic atomic constituents and to its microstructure.
The studies of pure magnetic phenomena on the nanosize scale
(domain wall structure, shape of domains and so on) can usually be
addressed with appropriate micromagnetic techniques (e.g., 
the finite element method). Quite extended regions (1--1000~nm and
up) can be studied with these techniques, but
such important ingredients of the problem as the atomic-scale
properties of defects and the microstructure are usually addressed
only phenomenologically. The first problem is related to the fact
that no first-principle calculations have been used to provide adequate
description of interatomic interactions on the atomistic scale, whereas the
second one is due to lack of adequate theoretical description of
the microstructure. This microstructure plays an especially important
role in the properties of hard magnets, because their high coercivity always
develops in quite characteristic nanoscale microstructures containing a high
density of such microstructural elements as grain boundaries, twins,
interphase and antiphase boundaries (APB), etc. Already here it is evident
that the theoretical description of the effect of microstructure on the
hysteresis loop of a hard magnet presents a rather complex task due to the
inherently multiscale nature of the problem and the presence of several
entirely different interactions. Interaction of domain walls (DW)
with many important defects is determined by variations of the microscopic
interaction parameters within the regions of atomistic size ($\sim $1~nm).
The DW width (5--10~nm in most hard magnets) is another length
scale, while the microstructure itself has one or more additional length
scales (typically within the 10-200~nm range). Each of these length scales
is physically important, and all of them must be linked together in order to
describe the magnetic properties consistently. Therefore, a simultaneous
inclusion of magnetic interactions on different scales (from 0.1 nm to 1000
nm) and corresponding elastic interactions (responsible for a given
microstructure) seems to be important for the consistent description of
the hysteretic phenomena.

Whereas the qualitative considerations above have been around for
a long time, to our knowledge, no theoretical methods combining the
atomistic, microstructural and micromagnetic parts of the multiscale
coercivity problem in real systems have been suggested.
Below we show how one can combine the first-principles, micromagnetic and
microstructural calculations and apply them to study the rich physics
of hysteretic phenomena in hard magnets of CoPt type and
demonstrate that coercivity has a natural multiscale character and must be
studied using an approach including all the relevant scales and interactions
on an equal footing.

Below we first discuss the choice of studied CoPt type systems
in section~\ref{secchoice}. Microstructure of these materials and its
theoretical description is described in section~\ref{secmicro}.
In section~\ref{secmfa} we outline the microscopic mean-field technique
replacing the continuous micromagnetic approach for the studies of
microscopically defined microstructures. The results on atomistic scale
obtained in \emph{ab initio} calculations will be discussed in
section~\ref{secpin} in connection with APB defects.
In section~\ref{secreal} we combine the description of
all the relevant length scales and interactions and find the
structure of a macrodomain wall in a realistic polytwinned
microstructure. Section~\ref{secconc} concludes the paper.

\section{\label{secchoice}Choice of systems}


Intermetallic hard magnets CoPt, FePt and FePd develop high
coercivities in the tetragonal L1$_{0}$ phase. 
All the accumulated knowledge about this magnet family suggests that it
can serve as a prototype for the problem we specified in introduction.
First of all, the main microstructural features in these alloys are well
established experimentally on the nanoscale level (10--100~nm) 
which is comparable with $\delta$. Notably, the microstructure is
dominated by coherent crystallographic defects (twins and
APB's~\cite{Soffa2000}) and not by grain boundaries, as in other hard
magnets. This circumstance provides two major advantages: the
microstructure may be consistently simulated theoretically (see section
~\ref{secmicro}), while
the microscopic properties of defects are tractable in modern
first-principles calculations. Indeed, the description of electronic
structure of alloys of 3d atoms with Pd and Pt is sufficiently reliable,
and the range of perturbation near an APB (1--5~nm) may also be covered.
Such parameters as magnetic anisotropy and effective exchange coupling
are also routinely computed. With this methodical background,
identification of the links
between the microstructure and magnetic properties of bulk CoPt-type magnets
is of interest both for the understanding of their own physics, and for the
general theory of hysteretic phenomena. Knowledge of such links may prove
useful in the design of future nanofabricated magnets.

\section{\label{secmicro}Microstructure:
Polytwinning and antiphase boundaries}

Typical thermal processing for CoPt type alloys involves 
high-temperature annealing in the disordered A1 (fcc) phase
area of the phase diagram followed by a rapid quench and
ageing at a lower temperature in the L1$_0$ phase area.
It is known from transmission electron microscopy (TEM)
experiments~\cite{Leroux91,Soffa2000} that the
microstructural evolution during ageing includes two
stages, the `tweed' stage, and the twinning stage.
The `tweed contrast' corresponds to a uniform pattern
formed by small (1--10~nm) ordered domains with a
dominant $\{101\}$ orientation of interfaces and essentially
random distribution of the $c$-axis directions. When the
growing ordered domains reach some characteristic size
(10-20~nm), a new pattern develops, with the formation
of large `polytwinned' stacks containing ordered `bands'
(`$c$-domains') with two alternating directions of the
$c$-axis making $\pi/2$ angles with each other.
The interfaces between the $c$-domains within a
polytwinned stack are all parallel to each other and
lie in one of the {101} planes. The main features of
microstructure on both these stages stem from the presence
of three possible orientations of the $c$-axis of the
ordered phase relative to the cubic disordered phase.

Physically, polytwinning is explained by the fact that
such structures eliminate the volume-dependent part of
the elastic energy stemming from the coexistence of
ordered domains with different $c$-axis
directions~\cite{Roitburd,Kh-Sh}. Main features of
the microstructural evolution at tweed and polytinning
stages were reproduced using a phenomenological 2D
model~\cite{Khach}. Recently a microscopic model 
for the description of elastic interactions in
alloys undergoing L1$_0$ type ordering was proposed
and used~\cite{BPSV} in detailed studies of various
aspects of microstructural evolution. In such alloys
the elastic interaction turns out to be effectively
non-pairwise, but it retains the standard long-wavelength
`elastic singularity' of the $k^{-2}$ type~\cite{BPSV}.
Therefore, the elastic interaction, as the dipole-dipole
interaction, formally has an infinite range. However,
due to the fact that the elastic energy of a
non-polytwinned array of ordered domains grows faster
with the domain size compared to the contribution from
the interface tension, there is a characteristic size
$l_0$ at which the elastic interaction begins to affect
the microstructural evolution~\cite{BPSV}. This size
(usually about 10~nm) corresponds to the average size
of domains on the tweed stage; the thickness of twins
on the polytwinned stage can not be less than $l_0$.

As it was shown earlier~\cite{MDW}, the twin boundaries
strongly affect the structure of `macrodomain walls' (MDW)
crossing the polytwinned stacks by splitting them in
segments that are coupled only magnetostatically.
Each segment is pinned individually by pinning
centers within the $c$-domain. This leads us to 
another generic feature of the polytwinned stage~---
the $c$-domains are interspersed with antiphase boundaries
(APB) between ordered domains with the same direction of the
$c$-axis and an antiphase shift along this axis.
The APB's act as pinning centers for DW's
due to a local suppression of MCA
(see Section~\ref{secpin}). Clearly, the efficiency
of pinning depends on the pattern formed by APB's.
So far, very
little is known about the dependence of this pattern
on the alloy composition and thermal processing. For
example, in Fig.~5 of Ref.~\onlinecite{Leroux91}
the APB's in a CoPt sample appear to be rectilinear,
while the available TEM images~\cite{Soffa2000}
for FePd alloys reveal quite dense patterns of curved
APB's, albeit with some tendency to a preferred
alignment. The former case is readily reproduced
in microscopic simulations~\cite{BPSV}, while
the prerequisites for the latter pattern are
unclear. It is quite obvious that pinning efficiency
should strongly depend on the prevailing APB pattern,
and each case should be studied separately.
One way to understand the combined effect of
MDW splitting and pinning in various microstructures
is to take characteristic
microstructures obtained in microscopic simulations
and add the magnetic degrees of freedom using the
method described in section~\ref{secmfa}. Since
magnets operate at relatively low temperatures
where diffusion is suppressed, the microstructure
may be fixed in magnetic simulations. In this
paper we will illustrate this approach by
finding an equilibrium structure of MDW's
in a polytwinned microstructure
taken from a microscopic simulation where APB's
are predominantly rectilinear.

\section{\label{secmfa}Magnetic simulations: Microscopic mean-field method
or micromagnetism}

The simplest way to add the magnetic degrees of freedom to the microscopically
simulated microstructure is to use the atomistic version of the micromagnetic
approach which turns out to be convenient in our problem where the relevant
length scales are of the order of 5--50~nm.

Consider a binary alloy AB with the classical Hamiltonian 
\begin{eqnarray}
H=H_{\mathrm{conf}}\{n_i\}+\sum_{i<j}n_{i}n_{j}\left[ -J_{ij}{\vec{\mu}}_{i}{%
\vec{\mu}}_{j}+{\vec{\mu}}_{i}\widehat{D}_{ij}{\vec{\mu}}_{j}\right] 
\nonumber \\
+\sum_{i}n_{i}[\epsilon_i({\vec{\mu}}_{i})-\mathbf{H}_{0}{\vec{\mu}}_{i}]
\label{H-Heisenberg}
\end{eqnarray}
where $H_{\mathrm{conf}}$ is the configurational part of the Hamiltonian; $i$
and $j$ run over lattice sites; $n_{i}=1$ if site $i$ is occupied by a
magnetic atom A and $n_{i}=0$ otherwise (let B be non-magnetic for
simplicity); ${\vec{\mu}}_{i}$ is the rigid classical magnetic moment of the
atom at site $i$; $J_{ij}$, the exchange parameters; $\mathbf{H}_{0}$, the
external magnetic field; $\epsilon_i({\vec{\mu}}_{i})$, the MCA energy equal
to $-b_{i}({\roarrow\mu}_{i}\mathbf{e}_i)^{2}$ for easy-axis anisotropy; and 
$\widehat{D}_{ij}$, the magnetic dipole-dipole interaction tensor.

Due to the fact that exchange interaction dominates at small distances, in
the vast majority of problems the magnetic moments may be considered to be
everywhere close to ferromagnetic alignment, and the angles between $\vec{\mu%
}_i$'s with non-vanishing $J_{ij}$ are usually relatively small. Therefore,
the exchange term in $H$ should be considered as the expansion of the total
energy in the angles of deviation from this alignment. The parameters $J_{ij}
$ are thus identified with the second derivatives of the total energy in
respect to these angles, rather than with the parameters of the Heisenberg
model defined to account for arbitrary angles between $\vec{\mu}_i$'s.

The inhomogeneous states of the system may be described by the free energy 
\begin{equation}
F=\langle H+T\ln P\rangle
\end{equation}
where $T$ is temperature, and $P$ is the distribution function. For the
canonical statistical ensemble $\ln P_0=\beta(F-H)$, and the system is
homogeneous. For inhomogeneous magnetic states with a fixed atomic
configuration, let us introduce the ``generalized Gibbs distribution'' $P=%
\exp[\beta(\widetilde F-H_{\mathrm{eff}})]$ as it was done~\cite{GGD} for $%
H_{\mathrm{conf}}$. In the theory of second-order phase transitions the
effective Hamiltonian $H_{\mathrm{eff}}$ is a functional of the macroscopic
order parameter field~\cite{LL5}, while in our microscopic case it contains
some `effective parameters' depending on the environment, i.e. on the
average values of ${\vec\mu}_i$ and, if necessary, of their powers and
correlators. In the simplest mean-field approximation (MFA) which will be
used in this paper, in equilibrium these parameters are the `mean fields' $%
\mathbf{H}_i=\mathbf{H}_0+\sum_j(J_{ij}-\widehat D_{ij})c_j\mathbf{m}_j$
where $\mathbf{m}_j=\langle{\vec\mu}_i\rangle$ and $c_j=\langle n_i\rangle$
are the local magnetizations and concentrations. Specifically, in this
approximation we have 
\begin{equation}
H_{\mathrm{eff}}=\sum_i n_i\left[-\mathbf{H}_i{\vec{\mu}}_i +\epsilon_i({%
\vec{\mu}}_i)\right],
\end{equation}
and the free energy 
\begin{equation}
F=\langle H-H_{\mathrm{eff}}\rangle -T\ln\mathrm{Tr}\exp(-\beta H_{\mathrm{%
eff}})  \label{genGibbs}
\end{equation}
after averaging assumes its final form 
\begin{equation}
F=-E_{J,DD}-T\sum_{i}c_{i}\ln \int d\widehat{\mu }_{i}\exp \left[\beta(%
\mathbf{H}_{i}{\vec{\mu}}_{i}-\epsilon_i)\right].  \label{F}
\end{equation}%
Here $E_{J,DD}=\sum c_{i}c_{j}\mathbf{m}_{i}(-J_{ij}+\widehat{D}_{ij})%
\mathbf{m}_{j}$ is the total exchange and dipole-dipole energy. Equilibrium
states for given boundary conditions may be found by solving the
`self-consistency' relations $\mathbf{m}_{i}=-\partial F/\partial \mathbf{H}%
_{i}$.

Let us emphasize that the above form of $P$ is actually a \emph{reduced}
Gibbs distribution. The effective Hamiltonian contains effective parameters
depending on the averages over dynamical variables, while the Gibbs
distribution contains the Hamiltonian with actual interaction parameters. In
particular, the generalized Gibbs distribution with the effective
Hamiltonian and proper boundary conditions may be used to describe
inhomonegeous equilibrium states (e.g., the DW's), while the
standard Gibbs distribution may not, because it retains the full symmetry of
the Hamiltonian.

In cases when MFA is inadequate (e.g.~in frustrated systems) one can use
more refined statistical methods to calculate $F$, e.g.~the cluster
variation method~\cite{Finel} or the cluster field method~\cite{VS},
although this will considerably complicate the calculations.

The microstructure is defined by the set of $c_i$. For the studies of simple
configurations this set may be prepared `by hand'. More realistic
configurations may be obtained in microscopic simulations~\cite{BPSV}.

If magnetization $\mathbf{M}(\mathbf{r})$ slowly varies in space and is
constant in magnitude, Eq.~(\ref{F}) reduces to the micromagnetic free
energy~\cite{Aharoni}. In this case all choices of $J_{ij}$ and $b_{i}$ in
the defect-free regions are equivalent if they produce the same macroscopic
properties $A$ and $K$. However, variation of $J_{ij}$ and $b_{i}$ at
interatomic distances near defects like APB's must be studied using
first-principles techniques. MFA (or more refined) calculations with these
parameters may be used to describe DW interaction with the defect at the
length scale of $\delta $. At larger, microstructural length scales
micromagnetic methods \cite{Kr1996,Fid-Sch} may be used
with singularities of $A$ and $K$ at the defects. However, as
we will see below, in hard magnets the microscopic approach also
turns out to be convenient for the studies of regions containing
up to $\sim10^6$ atoms; in such calculations some model $J_{ij}$
and $b_i$ reproducing the actual defect properties may be used.

\section{\label{secpin}Atomistic scale: pinning at antiphase boundaries}

As it was noted above, the DFT methods are quite reliable for
$d$-systems like Co--Pt. These methods were not used to study the
effect of lattice imperfections on the hysteretic phenomena because
the extended range required to simulate a DW is certaily
inaccessible for any band structure technique.
However, for CoPt-type alloys such study is feasible both in a
direct \emph{ab initio} approach (large supercell modelling
of DW and defect) and on the model level (calculation of parameters
of Hamiltonian). In the present paper we are setting up a technique
compatible with microstructural simulations, so we will discuss only
the latter approach. Before we proceed with this calculation let us
first discuss the physics of hysteretic phenomena on the atomistic level.

As we noted above, polytwinned alloys usually contain a high density of
APB's in the $c$-domains; it was argued that pinning on them may explain
high coercivity~\cite{Shur-CoPt,Zhang94}. In this section we will show using
first-principles calculations that pinning at APB's (developing on atomistic
scale alone) is indeed an important source of coercivity in given alloys.

If pinning is generated by planar faults like APB's, the maximum unpinning
threshold $H_{u}$ is achieved when DW segments (DWS) are parallel to the faults. If the
planar fault is represented micromagnetically as a slab of thickness $w\ll
\delta $ with modified exchange and anisotropy constants $A^{\prime }$ and
$K^{\prime }$, then $H_{u}$ is given by~\cite{pinning}: 
\begin{equation}
\frac{H_{u}}{H_{a}}\simeq \alpha \frac{w}{\delta }\left( \frac{A}{A^{\prime }%
}-\frac{K^{\prime }}{K}\right)   \label{Hu}
\end{equation}%
with $\alpha =\pi /3^{3/2}\simeq 0.60$ (our definition of $\delta $ contains
an extra factor $\pi $ compared to the notation of Ref.~\onlinecite{pinning}%
). Positive or negative $H_{u}$ corresponds to DW attraction to or
repulsion from the fault.

One should bear in mind that formula (\ref{Hu}) may be used only if the
range of exchange interaction (distance at which $A$ converges) is smaller
than $\delta$; otherwise, the representation of the planar fault by any
variation of $A$ across the fault is meaningless. In fact, this is the case
in all hard magnets, because $A$ is proportional to the sum $\sum_jJ_{ij}(%
\mathbf{r}_{ij}\mathbf{n})^2$ ($\mathbf{n}$ is the unit vector normal to the
APB) which converges very slowly~\cite{AHS}. Contrary to the homogeneous
case, $A(x)$ can not be conveniently computed using a summation in the
reciprocal space ($x\equiv\mathbf{rn}$ is the coordinate normal to the APB).
Therefore, one should find the `exchange contribution' to $H_u$ directly
using the microscopic representation of the defect and the formalism of
Section~\ref{secmfa}. Note that the problem of slow convergence occurs only
in the continuous formulation, because the above second-moment sum applies
only to the region where the DW profile $\mathbf{m}(x)$ may be replaced by
its second-order Taylor expansion. The parameters $J_{ij}$ may be calculated
using the local spin density functional approach~\cite{exch} which gives the
second derivatives of the total energy over the deviations of a pair of
spins from the equilibrium ferromagnetic alignment, in accordance with our
definition of $J_{ij}$. Technically, $H_u$ may be found as the maximum field
at which an equilibrium state of the DW at the APB is possible.

The modification of $K$ at the defect has a negligible effect on the DW
profile $\mathbf{m}(x)$ and hence may be accounted for perturbatively.
Therefore, we may divide $H_u$ in two parts, $H_u=H_{ue}+H_{ua}$, where $%
H_{ua}=\alpha H_a(w/\delta)(1-K^{\prime}/K)$, and $H_{ue}$ must be found
microscopically as described above, assuming that MCA is not affected by the
defect.

It is convenient to rewrite $H_{ua}$ in a different form more suitable for
first-principles calculation. Due to the fact that the MCA modification at
the defect may be treated perturbatively, the above formula for $H_{ua}$
does not depend on \emph{how} MCA behaves close to the defect, as long as
the perturbation falls off at distances smaller than $\delta$. In the `slab'
model used to derive (\ref{Hu}) the product $\Delta f_a=w(K-K^{\prime})$ may
be identified as the `anisotropy deficit' per unit area of the defect, i.e.
the difference in the free energies of the two large samples without and
with the defect. Thus, we obtain 
\begin{equation}
H_{ua}=\alpha H_a\Delta f_a/K\delta.  \label{Hua}
\end{equation}

The value $\Delta \epsilon _{a}$ of $\Delta f_{a}$ at $T=0$ may be directly
found using the first-principles calculations. To this end, it is necessary
to take a sufficiently large supercell containing the defect, calculate its
MCA energy, and subtract the corresponding MCA energy of the pure material.
In this context, `sufficiently large' means that $\Delta \epsilon _{a}$ is
converged with the supercell size. One should bear in mind that symmetry is
lowered at the defect, and MCA modification at the defect may lack the
easy-axis symmetry of the pure material.

We studied the modification of exchange and MCA at an isolated
(101)-oriented APB in CoPt, FePt and FePd using the tight-binding
linear-muffin-tin-orbital (TB-LMTO) method. We used elongated supercells of
up to 28 atoms with an APB in the middle (the geometry of the (101) APB
allows one to introduce a single APB per unit cell of $8n+4$ atoms).

The analysis of the exchange parameters for CoPt in such supercells in
conjunction with MFA simulations shows that changes in the local geometry
together with the modification of $J_{ij}$ at such APB generates only a weak
DW repulsion from the APB with $|H_{ue}|<1$~kOe. The total exchange
interaction between the pairs of atomic layers parallel to the (101) APB is
notably affected only in the close vicinity of the APB, particularly for the
next-nearest layers across the APB where extra nearest neighbors appear
compared to the pure material.

From the other hand, we found that MCA is strongly suppressed at the (101)
APB leading to a considerable DW \textit{attraction} to it. The values of $%
\Delta \epsilon _{a}$ per area of the APB corresponding to one formula unit
in each atomic layer are $5.0E_{a}$ for CoPt, $2.8E_{a}$ for FePt and $%
4.8E_{a}$ for FePd, where $E_{a}$ is the bulk MCA energy per formula unit.
Since high MCA is associated with L1$_{0}$ ordering, it is natural that
local disorder introduced by an APB suppresses MCA; one may expect it to
be a generic feature of APB's and other defects.

At low temperatures compared to the ordering transition the order parameters
and atomic structure of defects are almost independent on temperature. In
this temperature range (going up and beyond the room temperature) it is
reasonable to expect that $\Delta f_a$ and $K$ depend on $T$ in a similar
way. Therefore, an accurate estimate of $H_{ua}$ may be obtained using (\ref%
{Hua}) with the zero-temperature values of $\Delta f_a$ and $K$ and actual
values of $H_a$ and $\delta$ at the given $T$ (the dependence of $\delta$ on 
$T$ is also quite weak, because both $A$ and $K$ decrease with increasing $T$%
). At higher temperatures the disorder at APB's may increase considerably,
leading to a different $T$ dependence of $\Delta f_a$.

Using (\ref{Hua}) and the values of $H_{a}$ and $\delta $ from Ref.~%
\onlinecite{Kandaurova-JMMM} we obtain the threshold fields of 6.8, 3.7 and
1.2~kOe for CoPt, FePt and FePd, respectively; these values are consistent
with observed coercivities. Disorder at an APB (which is always present in
real alloys with non-nearest-neighbour configurational interaction~\cite%
{BPSV}) notably increases $\Delta \epsilon _{a}$, and hence $H_{ua}$. This
effect should also be expected at high temperatures affecting the
temperature dependence of the coercive force.

\section{\label{secreal}Macrodomain walls in a
realistic microstructure}

Although experimental observations are too scarce for reliable conclusions,
in actual microstructures the APB's are rarely rectilinear, except for
advanced stages of annealing (like the sample shown in Fig.~5 of Ref.~%
\onlinecite{Leroux91}); at early stages of annealing the $c$-domains contain
a large density of intricately curved APB's~\cite{Soffa2000}. Therefore,
$H_{u}$ in real alloys may be significantly lower compared to the values
obtained above for the plane-parallel configuration of DW's and APB's. The
best coercive properties should be achieved as a result of the competition
of a number of factors depending on the annealing time, most notably
$H_{u}$, $H_{s}$, $l_{d}$ and $l_{m}$. A complicated interplay of these
factors may
be studied by direct simulation of MR in realistic microstructures obtained
in microscopic simulations. To illustrate this approach, we found a stable
configuration of two MDW's in the CoPt model described above with a
microstructure similar to the experimental one shown in Fig.~5 of
Ref.~\onlinecite{Leroux91}. We took a simulated structure similar
to Fig.~5 of
Ref.~\onlinecite{BPSV} but obtained~\cite{PVunp} with a larger tetragonal
distortion $\varepsilon_m=0.2$, and augmented it to a 256x256x1 bct
simulation box ($\sim70$~nm) using the periodic boundary conditions. In order to reproduce
DW attraction to APB's in accordance with our band structure calculations,
we put $b_{i}=0$ in the close vicinity of APB's (defined by a certain
combination of local order parameters), and $b_{i}=b$ at all other sites
with easy axis $\mathbf{e}_i$ coinciding with the local tetragonal axis
(also found using local order parameters). The exchange parameters $J_{ij}$
were set as it was specified above, making the exchange constant $A$
continuous at (101) APB's. This choice of $b_i$ and $J_{ij}$ follows the
general prescription given above after Eq.~(\ref{F}); the unpinning
threshold for parallel DW and APB was made to be close to its value
obtained in Section~\ref{secpin} for CoPt.

\begin{figure}[tbp]
\epsfig{file=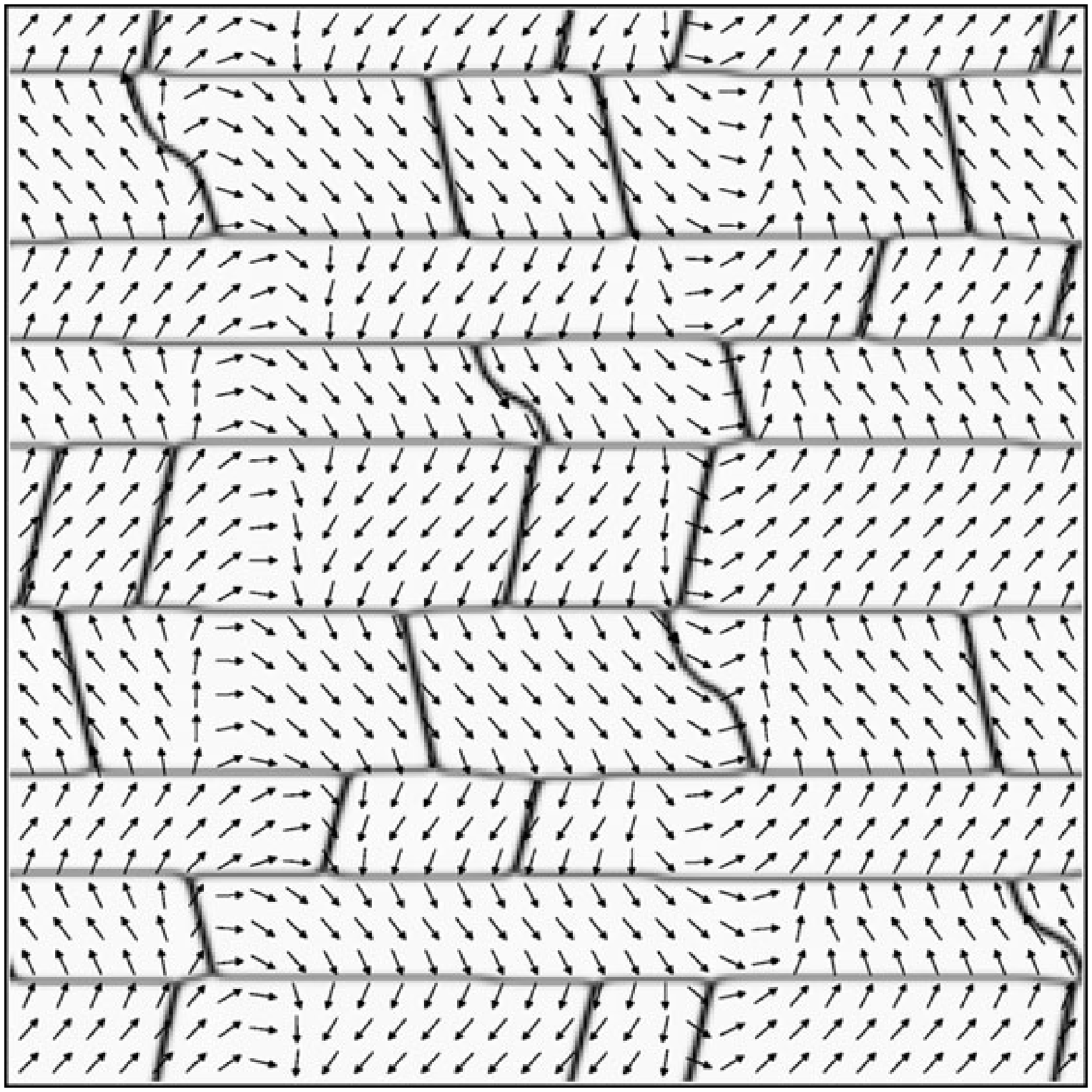,width=0.4\textwidth}
\epsfig{file=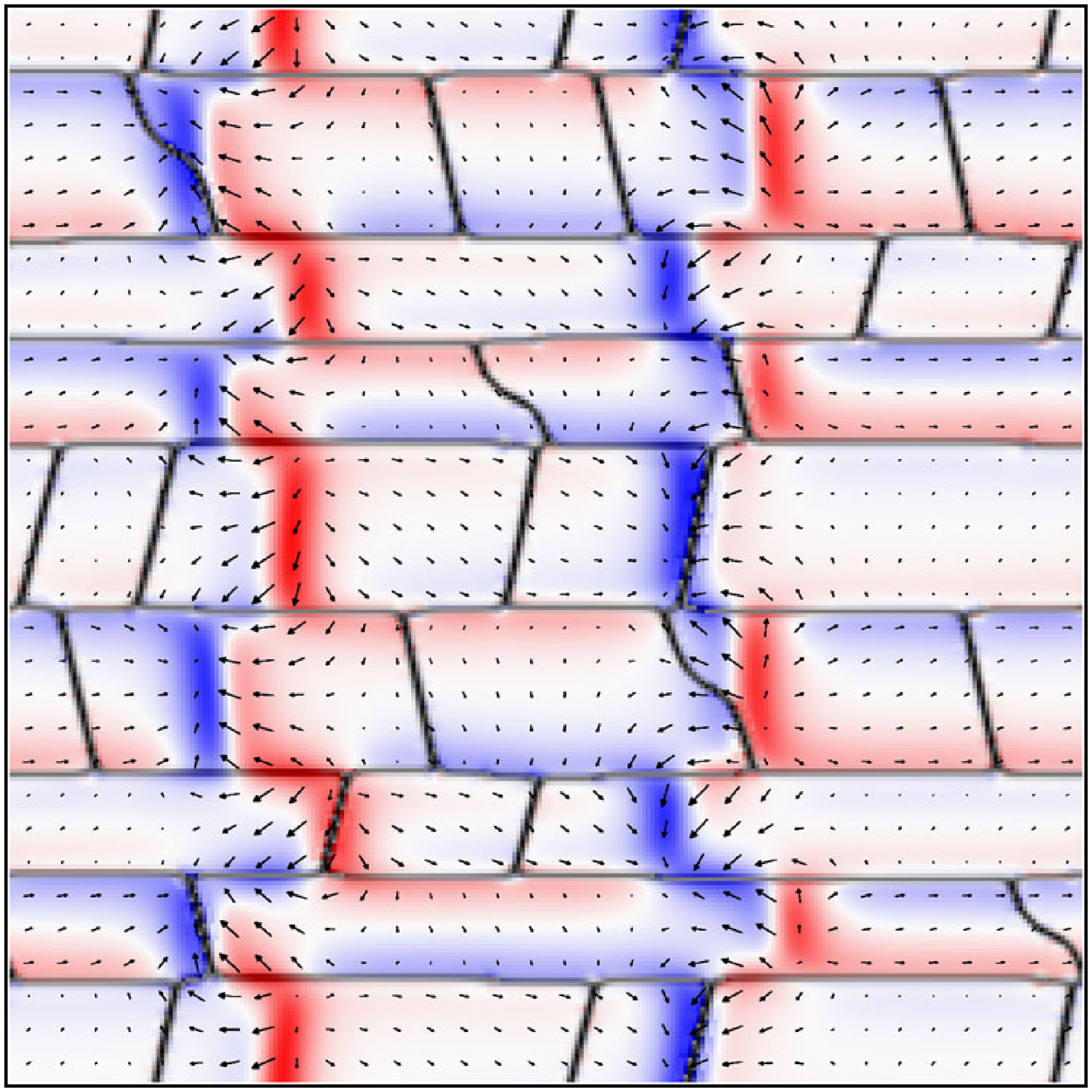,width=0.4\textwidth}
\caption{Macrodomain walls simulated for the model of a polytwinned CoPt
magnet (quasi-2D box with $\sim70$~nm edge).
The distributions of magnetization (top) and dipole fields (bottom)
are virtually planar and are shown by arrows; color shows the magnetic
charge density $\rho=-{\rm div}\,{\bf M}$ (red for positive, blue for
negative). The black lines show horizontal twin boundaries and APB's
within the $c$-domains.}
\label{real}
\end{figure}

The resulting magnetic structure is shown in Fig.~\ref{real} illustrating
and extending our conclusions about the properties of MDW. The MDW's are
seen to be `soft'; DWS's in the $c$-domains have a considerable freedom in
their relative displacement, and several of them are pinned at APB's.
Notably, the DWS's do not change their shape strongly when partially pinned
by APB's (e.g., in the upper-left corner). Strong magnetic fields `left
behind' by displaced DWS's are responsible for the attraction of DWS's in an
isolated MDW.

Two separate sources of coercivity in polytwinned CoPt type
magnets, domain wall pinning at antiphase boundaries and
splitting at twin boundaries~\cite{elsewhere}, are clearly
present in our simulations for a realistic microstructure.
Splitting may strongly affect the efficiency of DW pinning,
and it is important to account for both these effects
simultaneously in order to describe the coercivity.

\section{\label{secconc}Conclusion}

We have described the DW properties in polytwinned CoPt-type
systems where two sources of coercivity, domain wall pinning at
antiphase boundaries and splitting at twin boundaries, are
included simultaneously using a realistic microstructure.
Pinning is related to MCA lowering at the planar defect
(APB) whereas splitting comes from the competition between the
long-range dipole-dipole and short-range exchange interactions.

From the applied point of view, it is usually assumed that polytwinning
is detrimental for high coercivity. However, splitting may increase
the efficiency of DW pinning by allowing the DWS's to `find'
pinning sites independently.
It seems that the application potential of polytwinned microstructures
has not been exhausted, and better properties may be achieved on the basis
of this new theoretical understanding.

From the methodical point of view, we have shown that the coercivity is an
intrinsic multiscale property in hard magnets and to describe this property
one has to use a theory which necessary combines a synergy of different length
scales in a presence of physical interactions of entirely different nature.
Whereas the different aspect of the problem of magnetization reversal can be
studied using one level approach, the problem as a whole requires more
complicated description. We believe that such synergistic type of approach
represent a generic feature for any theory to be succesful in area of nanoscale
type of magnets: bulk systems, multilayers, clusters and nanowires.

We thank H. Kronm\"uller, R. Skomski and V.G. Vaks for useful discussions,
and members of the
DOE Nanomagnets CSP Group for their support and interest to this work.
This work was carried out at Ames Laboratory, which is operated for
the U.S. Department of Energy by Iowa State University under Contract
No. W-7405-82. This work was supported by the Director for Energy
Research, Office of Basic Energy Sciences of the U.S. Department of
Energy.

\end{document}